\documentclass[structabstract]{aa}  
\usepackage{graphicx}
\usepackage{txfonts}
\usepackage[authoryear]{natbib}
\bibpunct[; ]{(}{)}{;}{a}{}{,}
\begin{document}

\newcommand{\wise}{WISE\,J0720}

\title{Properties of the solar neighbor WISE\,J072003.20$-$084651.2
\thanks{Based on observations made with the Southern African Large
Telescope (SALT), the DDT Proposal Code \mbox{2013-2-RSA-016}.}}

\author{
V.~D.~Ivanov\inst{1}
\and
P.~Vaisanen\inst{2,3}
\and
A.~Y.~Kniazev\inst{2,3,4}
\and
Y.~Beletsky\inst{5}
\and
E.~E.~Mamajek\inst{6}
\and
K.~ Mu\v{z}i\'{c}\inst{1}
\and
J.~C.~Beam\'in\inst{1,7}
\and
H.~M.~J.~Boffin\inst{1}
\and
D.~Pourbaix\inst{8}\fnmsep\thanks{Senior Research Associate, F.R.S.-FNRS, Belgium}
\and
P.~Gandhi\inst{9,10}
\and
A.~Gulbis\inst{2,3}
\and
L.~Monaco\inst{1}
\and
I.~Saviane\inst{1}
\and
R.~Kurtev\inst{11,12}
\and
D.~Mawet\inst{1}
\and
J.~Borissova\inst{11,12}
\and
D.~Minniti\inst{12,13}
}

\offprints{V.~D. Ivanov, \email{vivanov@eso.org}}

\institute{
European Southern Observatory, Ave. Alonso de Cordova 3107,
Casilla 19001, Santiago 19, Chile
\and
South African Astronomical Observatory, PO Box 9, 7935 
Observatory, Cape Town, South Africa
\and
Southern African Large Telescope Foundation, PO Box 9, 7935 
Observatory, Cape Town,  South Africa
\and
Sternberg Astronomical Institute, Lomonosov Moscow State 
University, Moscow, Russia
\and
Las Campanas Observatory, Carnegie Institution of Washington, 
Colina el Pino, Casilla 601 La Serena, Chile
\and
Department of Physics and Astronomy, University of Rochester, 
Rochester, NY 14627, USA
\and
Instituto de Astrof\'isica, Facultad de F\'isica, Pontificia 
Universidad Cat\'olica de Chile, Casilla 306, Santiago 22, 
Chile
\and
Institut d'Astronomie et d'Astrophysique, Universit\'e Libre de
Bruxelles (ULB), Belgium 
\and
Department of Physics, Durham University, South Road, 
Durham DH1 3LE, UK
\and
School of Physics \& Astronomy, University of Southampton, 
Highfield, Southampton SO17 1BJ, UK
\and
Instituto de F\'isica y Astronom\'ia, Universidad de Valparaiso, 
Av. Gran Breta\~na 1111, Playa Ancha, 5030, Casilla, Chile
\and
Millennium Institute of Astrophysics (MAS), Chile
\and
Departamento de Ciencias F\'isicas, Universidad Andres Bello, 
Av. Republica 252, 8370251, Santiago, Chile
}

\date{Received 2 November 1002 / Accepted 7 January 3003}

\abstract
{The severe crowding towards the Galactic plane suggests that the 
census of nearby stars in that direction may be incomplete. Recently, 
Scholz reported a new M9 object at an estimated distance 
$d$$\simeq$7\,pc (WISE J072003.20$-$084651.2; hereafter \wise) at 
Galactic latitude $b$=2.3$^{\circ}$.}
{Our goals are to determine the physical characteristics of \wise, 
its kinematic properties, and to address the question if it is a 
binary object, as suggested in the discovery paper.}
{Optical and infrared spectroscopy  from the Southern African Large 
Telescope and Magellan, respectively, and spectral energy distribution 
fitting were used to determine the spectral type of \wise. The 
measured radial velocity, proper motion and parallax yielded its 
Galactic velocities. We also investigated if \wise\ may show X-ray 
activity based on archival data.}
{Our spectra are consistent with spectral type L0$\pm$1. We find no 
evidence for binarity, apart for a minor 2$\sigma$ level difference 
in the radial velocities taken at two different epochs. The spatial 
velocity of \wise\ does not connect it to any known moving group, 
instead it places the object with high probability in the old thin 
disk or in the thick disk. The spectral energy distribution fit 
hints at excess in the 12 and 22\,$\mu$m WISE bands which may be due 
to a redder companion, but the same excess is visible in other late 
type objects, and it more likely implies a shortcoming of the models 
(e.g., issues with the effective wavelengths of the filters for 
these extremely cool objects, etc.) rather than a disk or redder 
companion. The optical spectrum shows some H$\alpha$ emission, 
indicative of stellar activity. Archival X-ray observations yield no 
detection.}
{\wise\ is a new member of the Solar neighbourhood, the third nearest 
L dwarf. Our data do not support the hypothesis of its binary nature. 
}

\authorrunning{V.\,D.~Ivanov et al.}
\titlerunning{Nearby Star WISE J072003.20$-$084651.2}
\maketitle

\section{Introduction}\label{sec:intro}

Recently, Scholz \citet{sch14} reported the discovery of a new nearby
M-type star, WISE\,J072003.20$-$084651.2 (hereafter, \wise), 
demonstrating yet again that our knowledge of the late-type stellar 
content of the Solar neighbourhood is incomplete. Other new nearby
(D$\leq$10pc), 
low-mass mass objects discovered in the solar neighborhood include 
those reported by Artigau et al. \citep{art10}, Lucas et al. 
\citep{luc10}, Scholz \citet{sch11}, Luhman \citep{luh13}, Luhman 
\citep{luh14}, Scholz et al. \citet{sch14b}, among others. Nearby 
examples of low-mass stellar and substellar objects are important 
targets for understanding stellar interiors near the hydrogen-burning 
limit \citep{Dieterich14}, the diversity of stellar/substellar 
atmospheres and testing atmospheric models \citep[e.g.][]{Leggett10}, 
initial mass function \citep[e.g.][]{Kirkpatrick11}, activity trends
\citep[e.g.][]{Berger10}, the multiplicity and formation of low-mass
objects \citep[e.g.][]{Faherty10, Dieterich12}, and providing bright
low-mass targets for planet surveys \citep[e.g.][]{Blake10}.

\begin{figure}
\resizebox{\hsize}{!}{\includegraphics{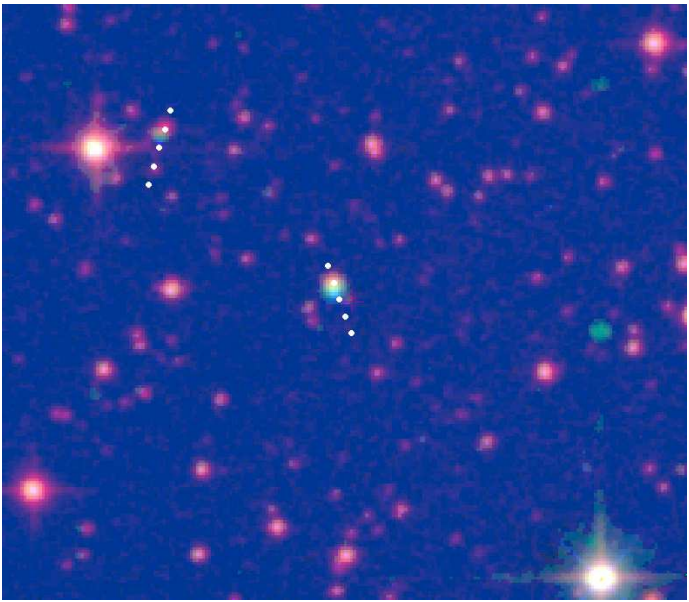}}
\caption{A false-color composite of DSSII IR (red), 2MASS $K_S$ (green), 
and our SofI $K_S$ image (blue) with $\sim$3.5$\times$3\,arcmin field 
of view, centered at \wise. North is up and East is to the left. Another 
HPM star 2MASS\,J07200708$-$0845589, not related to \wise, is visible to 
the North-East of \wise. The bright green dots on the West, with no 
counterparts on the other images, are cross-talk artifacts in the 2MASS.
The white dots show the positions of the two HPM stars in 50\,yr steps, 
over $-$50--150\,yr period, with respect to the DSS2 IR epoch (see 
Table\,\ref{table:new_HPM_star}).}\label{fig:image_PM}
\end{figure}

Typically, the missed objects are cool, making the Wide-field Infrared 
Survey Explorer \cite[WISE;][]{wri10} satellite an excellent tool for 
discovering them. Indeed, Scholz \citet{sch14} used color and proper 
motion criteria based on WISE and 2MASS \cite[Two Micron All Sky 
Survey;][]{scr06} photometry and astrometry to discover a new nearby 
late-M dwarf \wise\ -- a relatively bright, red object not far from 
the Galactic plane ($b$=+2.3\,deg; Fig.\,\ref{fig:image_PM}), where 
confusion has historically been a major obstacle for discovering new 
nearby, and high proper motion (HPM) objects 
\citep[i.e. Ivanov et al. ][]{iva13}. 
The star is cataloged in USNO B1.0 \citet{mon03} with proper motion
$\mu$$_\alpha$cos\,$\delta$=$-$36$\pm$13, 
$\mu$$_\delta$=$-$102$\pm$27\,mas\,yr$^{-1}$. Scholz \citet{sch14} 
astrometrically re-calibrated the archival images from the previous 
surveys that had detected \wise, and measured a trigonometric 
parallax of $\varpi$\,=\,142$\pm$38\,mas ($d$=7.0$\pm$1.9\,pc), 
and a revised proper motion of 
$\mu$$_\alpha$cos\,$\delta$=$-$41$\pm$2, 
$\mu$$_\delta$=$-$116$\pm$2\,mas\,yr$^{-1}$. The colors of the object 
were used to estimate a spectral type of M9$\pm$1. A comparison of 
the apparent mid-infrared (MIR) magnitudes of \wise\ with some M8-L0 
dwarf stars with well-known trigonometric parallax distances suggests 
a photometric distance in the range 6.0-8.4\,pc; a comparison of the 
IR photometry of the star to the best fit standard (M9V star 
LP\,944-20) hinted at a closer distance of 5.0$\pm$1.2 pc. The 
offset between the latter photometric distance and the trigonometric 
parallax distance prompted Scholz \citet{sch14} to propose that the 
target may be an unresolved binary. LP\,944$-$20 is considered as
young as, perhaps, 0.3\,Gyr \citep{Ribas03}. However, strong doubts 
about coevality of its purported moving group (Castor) by Mamajek et 
al. \citep{Mamajek13} suggest that the age of LP\,944$-$20 may not be 
as well constrained as previously thought. If LP\,944$-$20 is indeed 
young, or at least younger than \wise, then using it as an M9 template 
may partially explain the closer photometric distance to \wise.

Here we report comprehensive follow up observations of \wise\ aimed
at constraining its physical parameters.

\section{Observations}\label{sec:obs}

\subsection{Optical Spectroscopy}\label{sec:opt_spec}

We obtained long-slit spectra of \wise\ with the Robert Stobie 
Spectrograph \citep[RSS;][]{bur03,kob03} at the Southern African 
Large Telescope \citep[SALT;][]{buc06,odo06} in Sutherland, South 
Africa on November 13/14, 2013. Two spectroscopic setups were used, 
both with a 0.6\,arcsec wide slit at a parallactic angle 
(PA=$-$137.3\,deg) and with an on-chip $2\times2$ binning giving 
a spatial scale of 0.253\,arcsec\,pix$^{-1}$. The higher resolution 
spectrum using the PG1800 grating results in the spectral coverage 
of $\lambda$$\sim$7840--9000\,\AA\ and a resolution of 0.97\,\AA\ 
(0.33\,\AA\ per binned pixel), while the lower resolution setup 
using the PG0900 grating provided 2.19\,\AA\ (1.89\,\AA\ per binned 
pixel) resolution over $\lambda$$\sim$6300--9000\,\AA\ range. A 
900\,sec exposure was used for the former, and a 1060\,sec exposure 
for the latter. The seeing during both observations was 1.8\,arcsec. 
A Neon lamp arc spectrum and a set of calibration flats were taken 
immediately after the science frames. A spectrophotometric standard 
star, HR\,9087, was observed for both setups.

\begin{figure*}
\resizebox{0.5\hsize}{!}{\includegraphics{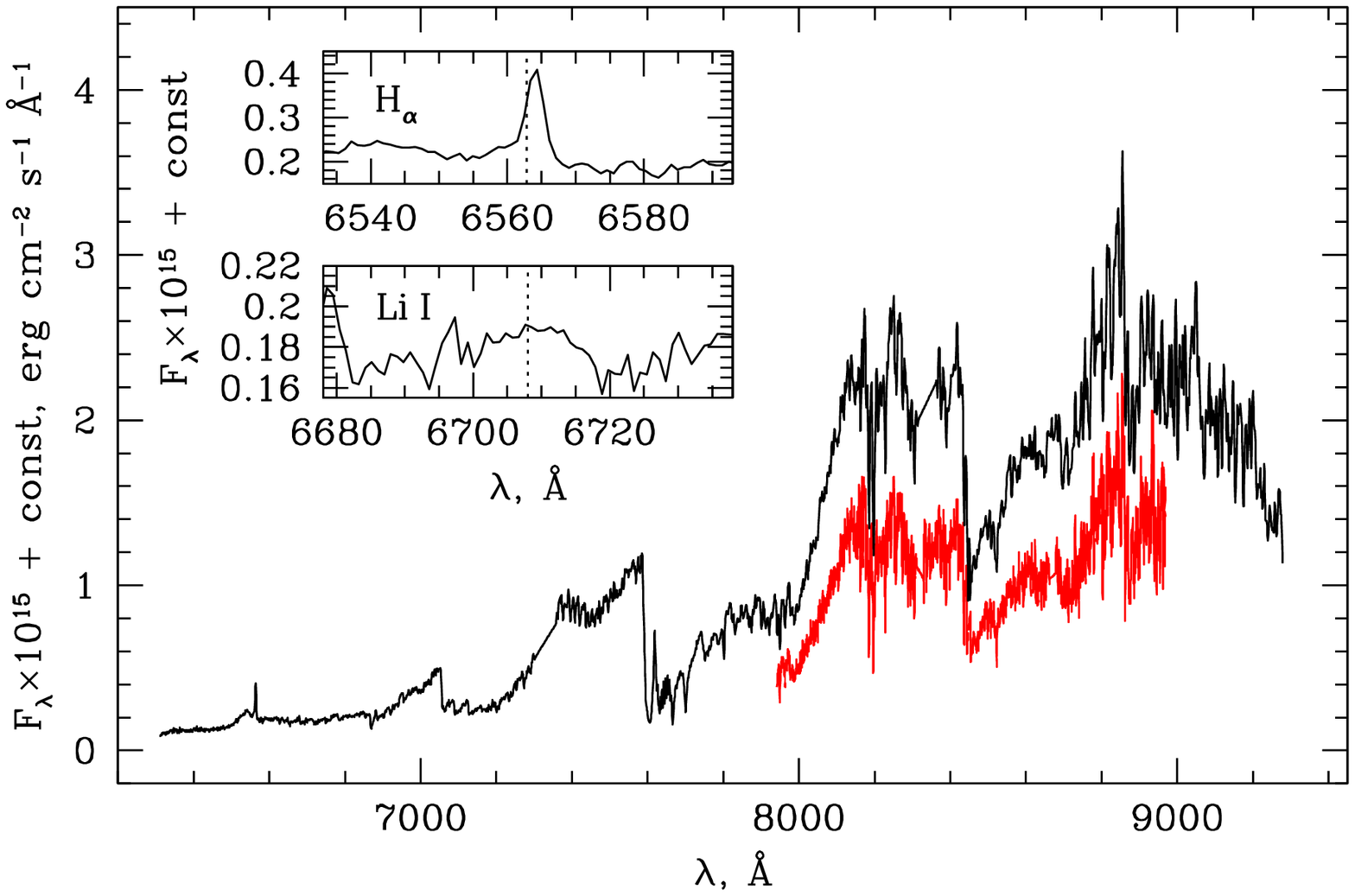}}
\resizebox{0.5\hsize}{!}{\includegraphics{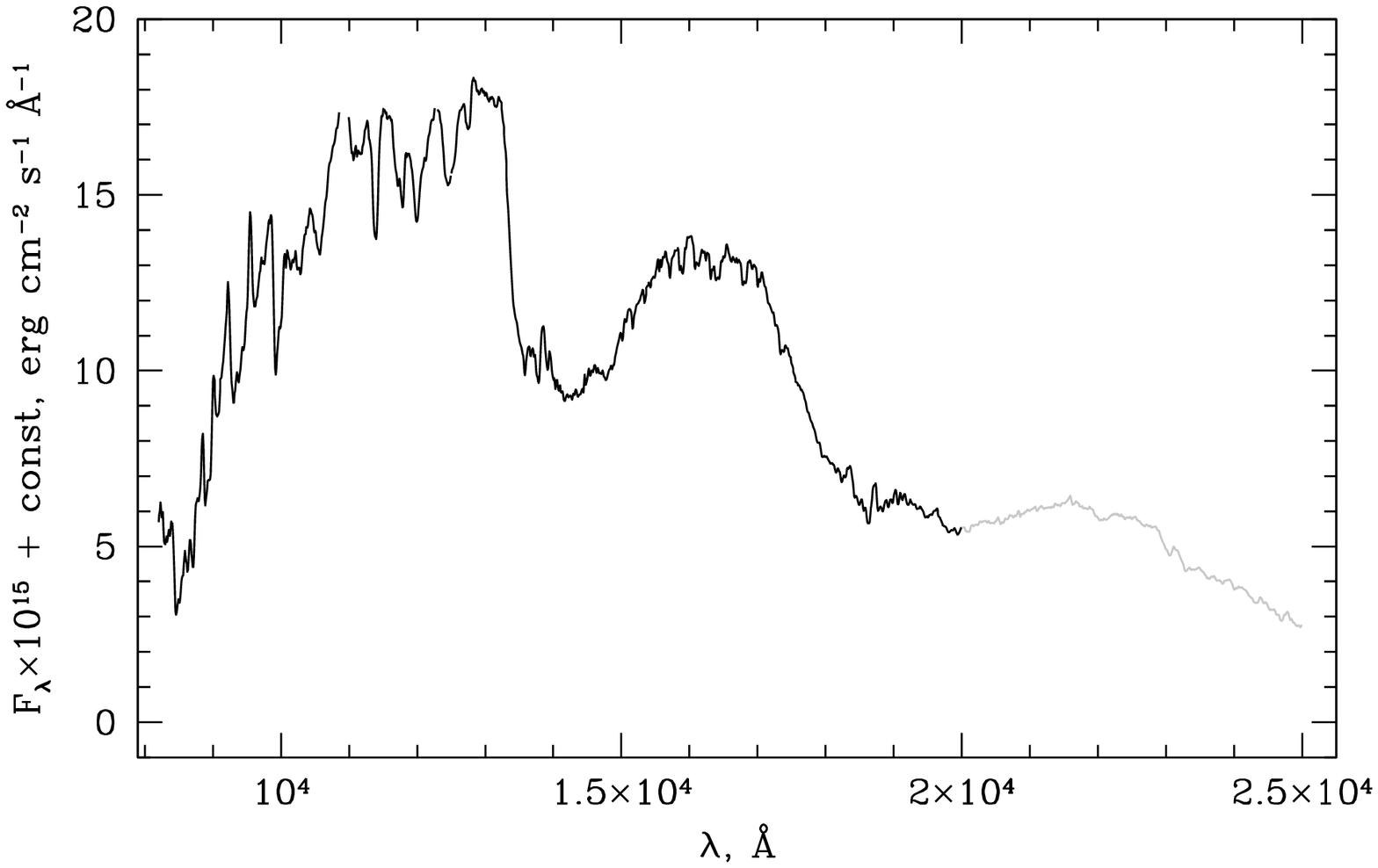}}
\caption{Optical (left) and NIR (right) spectra of \wise. The low
resolution optical spectrum is plotted in black, the medium-resolution
one in red. The redmost part of the NIR spectrum, affected by a 
wavelength calibration problem (see Sec.\,\ref{sec:nir_spec} for 
details) is shown in gray. The gap at $\sim$11000\,\AA\ is an omitted 
cosmic ray hit. The insets on the left zoom in on the regions around 
the H$\alpha$ and the Li\,I features, and the dotted vertical lines 
show their rest-frame positions. The mismatch between the fluxes of 
the three spectra are probably due to unaccounted for slit 
losses.}\label{fig:spectra_all}
\end{figure*}

The data were reduced following the same procedure as in Kniazev et 
al. \citet{kni13}. The SALT data pipeline PySALT \citep{cra10} 
products were used for primary reductions including the overscan, 
gain, cross-talk corrections, and mosaicing. Further routines in 
MIDAS\footnote{Munich Image Data Analysis System is distributed by 
ESO.} \citep{ban83,ban88,war92} and in the {\it twodspec} package 
in IRAF\footnote{IRAF is distributed by the NOAO, which is operated 
by the AURA under cooperative agreement with the NSF.} were used 
for wavelength calibration, frame rectification, and background 
subtraction of the 2D spectrum \citep{kni08}. The red part 
($>$8000\,\AA) of the spectra suffers from significant fringing 
effects, which are mitigated by using the calibration flats. The 
wavelength calibration for PG1800 and PG900 gratings resulted in an 
internal uncertainty of $\sigma$=0.03\,\AA\ and $\sigma$=0.13\,\AA\, 
respectively. The spectrophotometric standard was used for the 
spectral shape calibration, but no attempt was made to correct for 
slit losses or the changing pupil size of SALT. The velocities were 
corrected for heliocentric motion. The 1-dimensional (1D) spectra 
were extracted using a 5 pixel, $\sim$1.3\,arcsec, aperture. The 
final spectra are shown in Fig.\,\ref{fig:spectra_all} (left).

We measured equivalent widths of some of the more prominent features 
in our spectra with the IRAF task {splot}. The results are listed in 
Table\,\ref{table:eqw}. The errors reflect the uncertain continuum, 
and they are determined as the r.m.s. of multiple measurements. The 
Li\,{\rm I} line at $\lambda$=6707\,\AA was not detected, and we only 
give a 1$\sigma$ limit. The flux of the H$_\alpha$ emission line was 
7.6$\pm$ 0.2$\times$10$^{-16}$\,erg\,cm$^{-2}$\,s$^{-1}$. This error,
however, does not include uncertainties of absolute flux calibration
due to slit losses, that can easily amount to a factor of a few, as 
the comparison of the optical and the NIR spectra suggests.

The spectral indices of Kirkpatrick et al (1999) 
were also measured because of their utility for spectral typing. The 
results are listed in Table\,\ref{table:indices_Kirkpatrick}. The 
typical discrepancies between indices derived from the lower and the 
higher resolution spectra are 0.02--0.05, suggesting errors of that 
order.


\begin{table}
\caption{Equivalent widths for the more prominent spectral features 
of \wise\ in Angstroms. The line wavelengths $\lambda$ are listed in 
$\AA$.}\label{table:eqw} 
\begin{center}
\begin{tabular}{@{ }l@{ }l@{ }l@{ }l@{ }l@{ }l@{ }}
\hline\hline
Feature,\,$\lambda$          & Value             & Feature,\,$\lambda$ & Value          & Feature,\,$\lambda$ & Value       \\
\hline
\multicolumn{6}{c}{Optical features:} \\
H$\alpha$               6563 & $-$3.9$\pm$0.1~~  & Rb\,{\rm I}  7800   & 1.6$\pm$0.1~~  & Sc\,{\rm I}  8520   & 0.5$\pm$0.1 \\ 
Li\,{\rm I}             6707 & $\leq$0.07        & Rb\,{\rm I}  7948   & 1.1$\pm$0.1    &                     &             \\
\multicolumn{6}{c}{Near-Infrared features:}   \\                    
Na\,{\rm I}            11380 &    7.2$\pm$0.5    & K\,{\rm I}  12432   & 6.5$\pm$0.3    & Na\,{\rm I} 22062~  & 6.5$\pm$0.7 \\
Na\,{\rm I}            11400 &    6.8$\pm$0.3    & K\,{\rm I}  12527   & 6.7$\pm$0.9    & Ca\,{\rm I} 22640   & 0.9$\pm$0.2 \\
K\,{\rm I}+Fe\,{\rm I} 11690~&    6.4$\pm$0.2    & K\,{\rm I}  15168   & 3.6$\pm$0.2    &                           &             \\
K\,{\rm I}             11780 &    9.6$\pm$0.4    & Mg\,{\rm I} 15025   & 1.7$\pm$0.9    &                           &             \\

\hline
\end{tabular}
\end{center}
\end{table}

\begin{table}
\caption{Spectral indices for \wise, defined by 
Kirkpatrick et al (1999), and measured on our SALT spectra (the 
grating set up is given in brackets). The derived spectral types 
from the spectral sequences in the same paper (Figs.\,8--10) are 
listed.}\label{table:indices_Kirkpatrick} 
\begin{center}
\begin{tabular}{@{ }l@{ }l@{ }l@{ }l@{ }l@{ }l@{ }}
\hline\hline
Index~ & Value & Sp. Type~~~~~& Index~~ & Value & Sp. Type \\
\hline
Rb-a  & 1.15 (GR900)   & M7--L0 & Cs-b  & 1.16 (GR1800)  & L2--L4 \\
Rb-b  & 1.25 (GR1800)~~& L0--L2 & Cs-b  & 1.12 (GR900)   & L0--L3 \\
Rb-b  & 1.15 (GR900)   & M7--L0 & TiO-a & 1.82 (GR900)   & M9-L0 \\
Na-a  & 1.20 (GR1800)  & M9--L2 & TiO-b & 1.82 (GR1800)~~& M9-L1 \\
Na-a  & 1.19 (GR900)   & M9--L2 & TiO-b & 1.85 (GR900)   & M9-L1 \\
Na-b  & 1.35 (GR1800)  & M7--L0 & CrH-b & 1.04 (GR1800)  & M8-L2 \\
Na-b  & 1.33 (GR900)   & M7--L0 & CrH-b & 1.01 (GR900)   & M8-L2 \\
Cs-a  & 1.18 (GR1800)  & L1--L3 & FeH-a & 1.12 (GR1800)  & M7-9 \\
Cs-a  & 1.20 (GR900)   & L1--L3 & FeH-a & 1.12 (GR900)   & M7-9 \\
\hline
\end{tabular}
\end{center}
\end{table}

A radial velocity V$_{\rm obs}$=53.79$\pm$2.54\,km\,s$^{-1}$ was
measured from the higher resolution spectrum (taken at UT=00:42), 
using the absorption lines at $\lambda$=7947.60, 8521.13, 8183.255, 
and 8194.790\,\AA. The measurement of each line was verified with 
the nearest night sky lines using the method and programs described 
in Zasov et al. \citet{Zasov00}. The barycentric correction was 
22.77\,km\,s$^{-1}$, and the final heliocentric velocity was 
calculated to be V$_{\rm hel}$=76.56$\pm$2.54\,km\,s$^{-1}$. The 
final heliocentric velocity using H$\alpha$ emission line from the 
spectrum obtained with the PG900 grating was 
V$_{\rm hel}$=77.65$\pm$3.79\,km\,s$^{-1}$, fully consistent with 
the higher resolution spectrum.

\subsection{Near-Infrared Spectroscopy}\label{sec:nir_spec}

Low resolution long-slit NIR spectroscopy of \wise\ was carried out 
on November 16/17, 2013, with the Folded-Port InfraRed Echellette 
spectrograph \citep[FIRE;][]{sim08,sim13} at the Magellan telescope 
at Las Campanas Observatory. The instrument is equipped with a 
2048$\times$2048 HAWAII-2RG detector. We obtained four 1\,sec 
exposures in ABBA nodding pattern in the high-throughput prism mode, 
covering $\lambda$=0.82--2.51\,$\mu$m. The slit was 1\,arcsec wide, 
and the spatial scale was 0.15\,arcsec\,px$^{-1}$. The spectral 
resolution varies from R$\sim$500 in the J atmospheric window, to 
$\sim$300 in the K window. An A0V telluric HD\,65504 was observed to 
correct for the atmospheric absorption.

The data reduction performed was typical for the NIR spectra 
\citep[e.g. Ivanov et al. ][]{iva00,iva04}: sky subtraction, tracing 
and extraction of 1D spectra, wavelength calibration with Neon-Argon 
lamp spectra taken right after the target observation, combination 
of the 1D spectra in wavelength space, and telluric correction. We 
also multiplied the spectra by an A0 spectrum from the library of 
Pickles \citep{pic98} to remove the artificial features, introduced 
by the telluric itself. At this low resolution the arc lines above 
$\sim$20000\,$\AA$ were blended, affecting the wavelength calibration. 
Probably, the best way to treat this issue is to take two arcs - 
one with a narrow and another with a wide slit; then, to derive a 
wavelength calibration from the former, and to apply a shift to the 
latter using the lines in the wide slit arc that are not blended.
This problem resulted in smearing of the spectra in the $K$-band, so 
we did not use this part of the spectra in the subsequent analysis. 
The spectrum was pseudo-flux calibrated to the apparent 2MASS $H$ 
band magnitude. The other bands were not used because $J$ 
transmission is known to depend on the water vapor, and the $K$ 
spectrum not reliable, as pointed out above. The final spectrum is 
shown in Fig.\,\ref{fig:spectra_all} (right).

We measured the indices defined in McLean et al. \citep{McLean2003} 
because of their common use \citep{cus05} and straightforward 
parametrization versus spectral class over M- and L-type objects. The 
results are listed in Table\,\ref{table:indices_McLean}. The errors 
reflect only the formal uncertainties in the wavelength calibration, 
and were determined varying the position of the pass bands.

\begin{table}
\caption{Spectral indices for \wise, defined by McLean et al. 
\citet{McLean2003} and measured on our low-resolution FIRE spectrum. 
The derived spectral types from the spectral sequences in the same 
paper are listed. The Methane indices were omitted, because they 
cannot constrain the spectral type of M/L type objects, and the CO 
index was omitted because it is in an unreliable part of the 
spectrum.}\label{table:indices_McLean} 
\begin{center}
\begin{tabular}{@{ }l@{ }l@{ }l@{ }l@{ }l@{ }l@{ }}
\hline\hline
Index     & Value           & Sp. Type~~~~~~~~& Index   & Value           & Sp. Type \\
\hline
H$_2$OA~~~& 0.66$\pm$0.01~~~& L1$\pm$1.5 & H$_2$OD     & 1.02$\pm$0.01~~~& M9$\pm$1 \\
H$_2$OB   & 0.77$\pm$0.01   & L2$\pm$1   & $J$$-$FeH~~~& 0.90$\pm$0.01   & M8$\pm$2 \\
H$_2$OC   & 0.68$\pm$0.01   & L2$\pm$2   & $z$$-$FeH   & 0.76$\pm$0.04   & M7$\pm$2 \\
\hline
\end{tabular}
\end{center}
\end{table}

Medium resolution NIR spectrum of \wise\ was collected on November 
18/19, 2013 with the same instrument. We obtained four 120\,sec 
exposures in ABBA nodding pattern in Echelle mode, covering a similar 
range as above. The slit was 0.6\,arcsec wide. The spectral resolution, 
measured from ThAr lines, was R$\sim$5000. The A0V telluric star 
HD\,65504 was observed to correct for atmospheric absorption. We also 
obtained spectra of a K0IV radial velocity standard HD\,48381
\citep[V$_{\rm helio}$=40.722\,km\,s$^{-1}$, $\pm$0.0065 internal 
error, $\pm$0.0145 r.m.s. of 16 measurements;][]{Soubiran13}, and a 
corresponding A0 telluric star HIP\,32479. 

\begin{figure}
\resizebox{1.0\hsize}{!}{\includegraphics{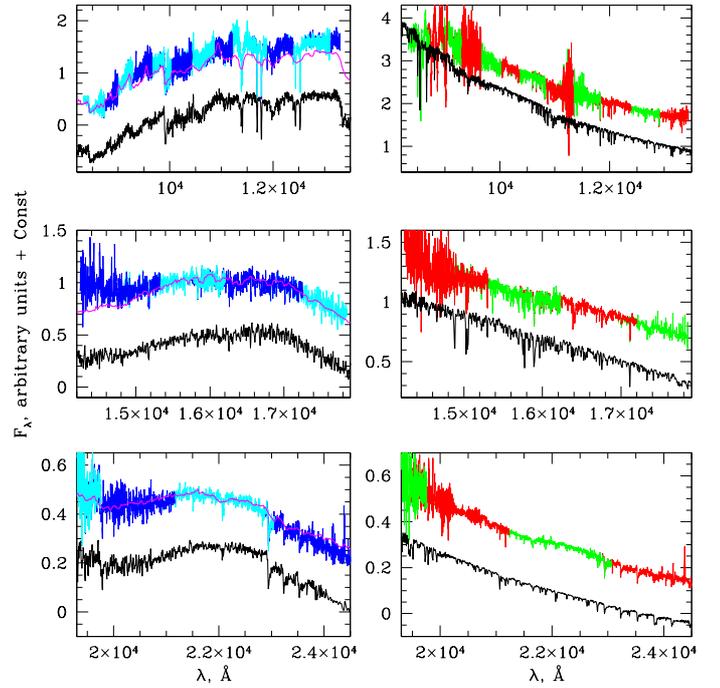}}
\caption{Medium-resolution NIR spectra of \wise\ (left) and the radial 
velocity standard HD\,48381 (right). The different orders are plotted
with alternating dark/light blue and red/green colors, respectively. 
The low resolution NIR spectrum of \wise\ is shown in magenta. For 
comparison we plot the M9V LP\,944$-$20 (left, black) and K0V 
HD\,145675 (right, black) spectra from the IRTF library 
\citet{cus05,ray09}, shifted down by 0.9 (on the top panels), 0.5 
(middle) and 0.3 (bottom). All spectra were normalized to unity for 
$\lambda$ between 15000 and 17000\,\AA.}\label{fig:spectra_HighRes}
\end{figure}

Each of the 21 orders was treated separately, and the reduction steps 
were the same as for the low resolution data. We used ThAr lamp 
spectra taken for the wavelength calibration of the bluest eight 
orders, and sky lines for the remaining thirteen. The final spectra 
are shown in Fig.\,\ref{fig:spectra_HighRes}. The global shapes of 
the low- and the medium-resolution NIR spectra do not match 
particularly well for $\lambda$$<$15000\,\AA. A number of reasons 
can account for that, e.g. a change of the atmospheric transmission 
between the observation of the target and the telluric calibrator, 
error in the spectral type of the telluric or wavelength dependent 
slit losses. The two spectra appear similar at $\lambda$$>$20000\,\AA,
however, despite the wavelength calibration problem mentioned above.

We measured the equivalent widths of some prominent features in the 
NIR spectrum, following the same procedure as for the optical spectra. 
The results are listed in Table\,\ref{table:eqw}. 

We used the IRAF task {\it fxcor} to measure the velocity difference 
between \wise\ and HD\,48381 with the CO band heads. They appear in 
two separate orders, so we obtained two independent measurements:
$\Delta$V$_{\rm obs}$=50.4$\pm$13.8 and 53.8$\pm$20.1\,km\,s$^{-1}$. 
We adopt throughout this paper their weighted average 
$\Delta$V$_{\rm obs}$=51.8$\pm$12.0\,km\,s$^{-1}$. The heliocentric 
correction for the radial velocity standard at the moment of the 
observation (UT=08:10) was $-$12.4\,km\,s$^{-1}$, yielding an 
observed radial velocity of 28.3\,km\,s$^{-1}$. Adding the measured 
difference we obtain an observed velocity of the target of 
81.1\,km\,s$^{-1}$. The heliocentric correction for the target at 
the moment of the observation (UT=07:41) was 21.4\,km\,s$^{-1}$, so 
the heliocentric radial velocity of the target was 
V$_{\rm hel}$=101.6$\pm$12.0\,km\,s$^{-1}$ (the implications from 
this measurement will be discussed in Sec.\,\ref{sec:kinematics}).

Finally, we tested the consistency of these results measuring the 
relative velocity difference of the CO band heads in the two orders, 
obtaining 11.0$\pm$7.9\,km\,s$^{-1}$ for the target, and 
14.6$\pm$10.3\,km\,s$^{-1}$ for the standard, respectively. The 
difference is bellow the measurement errors.

\subsection{Optical Imaging}

Optical $BVRi$ images of \wise\ were obtained with EFOSC2 
\citep{buz84,sno08} at the ESO New Technology Telescope on November 
15/16, 2013. It is equipped with a 2048$\times$2048 Loral/Lesser CCD, 
delivering 0.12\,arcsec\,px$^{-1}$ images over a 
$\sim$4.1$\times$4.1\,arcmin field. The following exposures were 
taken: 30, 20, 10, and 2\,sec in $BVRi$, respectively, with 2$\times$2 
binning for photometry, and ten 2\,sec images in $i$ with 2$\times$2 
binning for astrometry. 

The data reduction included bias subtraction and flat fielding. The 
photometric calibration was based on $BVRi$ observations of two 
standards: PG\,2213$-$006 and RU\,152. The magnitudes in the 
standard system m$_{\rm std}$ were calculated as
\begin{equation}
m_{\rm std} = ZP + m_{\rm inst} + 2.5 \times log_{10}(t) + c_1 \times sec\,z + c_2 \times colour
\end{equation}
where ZP was the zero point, m$_{\rm inst}$ was the magnitude in the 
instrumental system, c$_{\rm 1}$ was the extinction coefficient
(adopted from historic La Silla records), c$_{\rm 2}$ was the color 
coefficient; the values of the coefficients and the colors used in 
the color terms for each band are listed in Table\,\ref{table:phot}.

\begin{table}
\caption{Apparent magnitudes of \wise\ in the Vega system and calibration 
coefficients. NIR magnitudes were not derived because the core of the 
target was registered in the non-linear regime of the detector. We 
also list the literature data used for the spectral energy 
distribution fit: The Carlsberg survey \citep[][release 15; SDSS $r$ 
filter]{Evans02}, DENIS \citep{epc97}, 2MASS \citep{scr06}, and WISE 
\citep{wri10}.}\label{table:phot} 
\begin{center}
\begin{tabular}{@{ }c@{ }c@{ }c@{ }c@{ }c@{ }c@{ }}
\hline\hline
Band         & Zero point       & $c_1$ & $c_2$  &~$colour$~ & Apparent mag    \\
\hline
$B$          &~25.182$\pm$0.103~&~0.06~ &~0.0330~& $B$$-$$V$ &~20.778$\pm$0.140~\\
$V$          & 25.457$\pm$0.051 & 0.05  & 0.0730 & $B$$-$$V$ & 18.266$\pm$0.096 \\
$R$          & 25.648$\pm$0.029 & 0.04  & 0.0056 & $V$$-$$R$ & 15.838$\pm$0.031 \\
$i$          & 24.901$\pm$0.035 & 0.00  & 0.0760 & $R$$-$$i$ & 14.019$\pm$0.035 \\
CMC15\,$r$   &                  &       &        &           & 16.850$\pm$---   \\
DENIS\,$I$   &                  &       &        &           & 13.805$\pm$0.02  \\
DENIS\,$J$   &                  &       &        &           & 10.674$\pm$0.06  \\
DENIS\,$K_S$ &                  &       &        &           & 9.399$\pm$0.09   \\
2MASS\,$J$   &                  &       &        &           & 10.628$\pm$0.023 \\
2MASS\,$H$   &                  &       &        &           & 9.919$\pm$0.024  \\
2MASS\,$K_S$ &                  &       &        &           & 9.467$\pm$0.019  \\
WISE\,W1     &                  &       &        &           & 9.174$\pm$0.023  \\
WISE\,W2     &                  &       &        &           & 8.860$\pm$0.022  \\
WISE\,W3     &                  &       &        &           & 8.333$\pm$0.024  \\
WISE\,W4     &                  &       &        &           & 7.934$\pm$0.212  \\
\hline
\end{tabular}
\end{center}
\end{table}

\subsection{Near-Infrared Imaging}

Near-infrared (NIR) $JHK_S$ images of \wise\ were obtained with SofI 
\citep{moo98} at the ESO New Technology Telescope on November 13/14,
2013. The instrument is equipped with a 1024$\times$1024 Hawaii HgCdTe 
detector, delivering 0.29\,arcsec\,px$^{-1}$ images over a 
$\sim$4.9$\times$4.9\,arcmin field. Four jittered images were taken in 
each filter, and each image was the average of five 2\,sec frames in 
$J$, and five 1.182\,sec frames in $H$ and $K_S$, the total 
integration times in the three NIR bands are 40, 23.64, and 
23.64\,sec, respectively. The core of the object is at the non-linear 
regime even at the shortest detector integration time, so these images 
are used only for astrometry of the target, and to search for fainter 
nearby companions, particularly when future high angular resolution 
observations become available.

The data reduction includes the usual steps: flat fielding, sky 
subtraction, image alignment and combination. The photometric 
calibration is based on isolated non-saturated 2MASS stars in the 
field, spanning $J$$-$$K_S$$\sim$0.3-1.0\,mag range. The zero points 
(listed in Table\,\ref{table:phot}) were derived from 18, 15, and 19 
stars, respectively for $J$, $H$, and $K_S$. No significant color 
terms were found.

The astrometric calibration was based on all stars in the field, and 
yields the following position: $\alpha$=+108.7242\,deg=07:14:53.8, 
$\delta$=$-$8.780546\,deg=$-$08:46:49.97 (J2000), with positional 
errors of 0.1\,arcsec. A false-color composite of DSSII IR, 2MASS 
$K_S$, and our SofI $K_S$ image is shown in Fig.\,\ref{fig:image_PM}.

\subsection{X-ray Observations}

The {\em Swift} satellite \citep{Gehrels04} observed the field of 
\wise\ on June 04, 2004 starting UT=21:35. A full band X-ray 
Telescope (XRT) image was extracted (in Photon counting mode, 
OBSID=00049016001, and exposure time 529\,sec) from the UK Swift 
Science Data Centre, which is a pre-reduced product using standard 
procedures.

With only one source count being detected over the full energy range 
around the position of \wise, the object is undetected in X-rays. 
The 3-$\sigma$ uncertainty on the count rate is 
5.7$\times$10$^{-3}$\,ct\,s$^{-1}$. Assuming an optically-thin 
thermal plasma with temperature $kT$=0.3\,keV 
\citep[e.g.][]{Tsuboi03} implies a 0.3--2\,keV flux limit of 
1.2$\times$10$^{-13}$\,erg\,s$^{-1}$ cm$^{-2}$, or X-ray luminosity 
upper-limit of 5.3($\pm$2.9)$\times$10$^{26}$\,erg\,s$^{-1}$ for a 
distance of 6.07\,pc (see Sec.\,\ref{sec:kinematics}). The column 
density of obscuring gas along the line-of-sight to the source is 
assumed to be negligible, given the proximity of \wise.

The spectral energy distribution (SED) fit of \wise\ 
(Sec.\,\ref{sec:sed}) predicts a bolometric luminosity of 
2.02$\times$10$^{-4}$\,L$_\odot$, yielding 
$log$($L_X$/$L_{\rm bol}$)$\leq$$-$3.2.
This luminosity constraint is weaker than those available for other 
well-studied late-type dwarfs, and much weaker than for the nearest 
known brown dwarf (hereafter, BD) Luhman-16 \citep{Gandhi13}. It lies 
close to the level of saturated emission, 
$log$($L_X$/$L_{\rm bol}$)=$-$3, expected for the M9 spectral class 
\citep{Berger10}.

\section{Results}\label{sec:results}

\subsection{Kinematics of \wise}\label{sec:kinematics}

Weight-averaging the optical and the NIR heliocentric velocities 
(V$_{\rm hel}$=76.56$\pm$2.54 and 101.6$\pm$12.0\,km\,s$^{-1}$, 
respectively), we obtain 77.6$\pm$2.5\,km\,s$^{-1}$.
Combining our new position of \wise\ with the previous ones, 
reported by Scholz \citet{sch14}, we obtain improved proper motions 
of $\mu$$_\alpha$cos\,$\delta$=$-$39.1$\pm$2.1
$\mu$$_\delta$=$-$111.8$\pm$1.9\,mas\,yr$^{-1}$ and a parallax 
distance of 6.07\,pc, with 1-$\sigma$ interval between 5.12 and 
7.43\,pc. The transverse velocity is 
$V_{tan}$=3.4$\pm$0.1\,km\,s$^{-1}$.

We calculate the barycentric velocity of \wise\ on the Galactic
coordinate system using the proper motion and parallax from Scholz
\cite{sch14}, and our optical radial velocity (because of the smaller
uncertainty): ($U$, $V$, $W$)\,=\,$-$63.7$\pm$2.1, $-$65.4$\pm$1.4,
0.6$\pm$1.1\,km\,s$^{-1}$. The total barycentric speed is 
$S$=91.4$\pm$1.8 km\,s$^{-1}$. Adopting the Local Standard of 
Rest (LSR) from Schoenrich et al. 
\citet{Schoenrich10}, the velocity translates to LSR velocity 
(U$_{\rm LSR}$, V$_{\rm LSR}$, W$_{\rm LSR}$)\,=\,$-$52.6, $-$53.2, 
7.8 km\,s$^{-1}$, and total LSR peculiar motion of 75.2 km\,s$^{-1}$. 
We estimate kinematic membership probabilities of 65.1\%, 34.6\%, 
and 0.3\%\, to the thin disk, thick disk, and halo, respectively, 
using the velocity moments and the local density normalizations of 
the these dominant Galactic kinematic populations from Bensby et al. 
\citet{Bensby03}, and adopting the LSR velocity from Schoenrich et al. 
\citet{Schoenrich10}. Given the large LSR velocity (75\,km\,s$^{-1}$), 
and the probabilities calculated above, it seems most likely that 
\wise\ is either an old thin disk or a thick disk star. Comparing the 
velocity of \wise\, to field F and G dwarfs in Fig. 1 of Bensby et al. 
\citet{Bensby07} shows that it is in the velocity region overlapping 
thin and thick disk stars, and that if it belongs to the thin disk, 
it could be associated with the Hercules dynamical stream 
(U$_{\rm LSR}$, V$_{\rm LSR}$ $\approx$ $-$40, $-$50 km\,s$^{-1}$), 
which unfortunately is of negligible help in constraining its age. 
Using an epicycle orbit code\footnote{We follow the epicycle equations 
of Fuchs et al. \citep{Fuchs06}, and adopt Oort A and B parameters 
from Feast \& Whitelock \citet{Feast97}, local disk density from 
\citet{vanLeeuwen07}, LSR velocity from Sch\"onrich et al. 
\citet{Schoenrich10}, solar Galactocentric radius 8.0\,kpc, solar 
distance above Galactic plane of 20\,pc.} we estimate that the
Galactic orbit of \wise\ has perigalacticon $R_p$$\simeq$3.27\,kpc,
apogalacticon $R_a$$\simeq$8.44\,kpc, and it is highly eccentric:
$e$$\simeq$0.44.

A search of the XHIP catalog \citep{Anderson12}
finds only one Hipparcos star with velocity within 4\,km\,s$^{-1}$ 
of \wise: HIP\,96426 -- a slightly supersolar 
\citep[\protect{[Fe/H]=0.08;}][]{Boeche11} K dwarf at 
$d$$\simeq$46$\pm$4\,pc, and while sharing the velocity of 
\wise, it is too distant to potentially be a physical companion.

\subsection{Spectral Typing}\label{sec:sptyping}

\begin{figure}
\resizebox{\hsize}{!}{\includegraphics{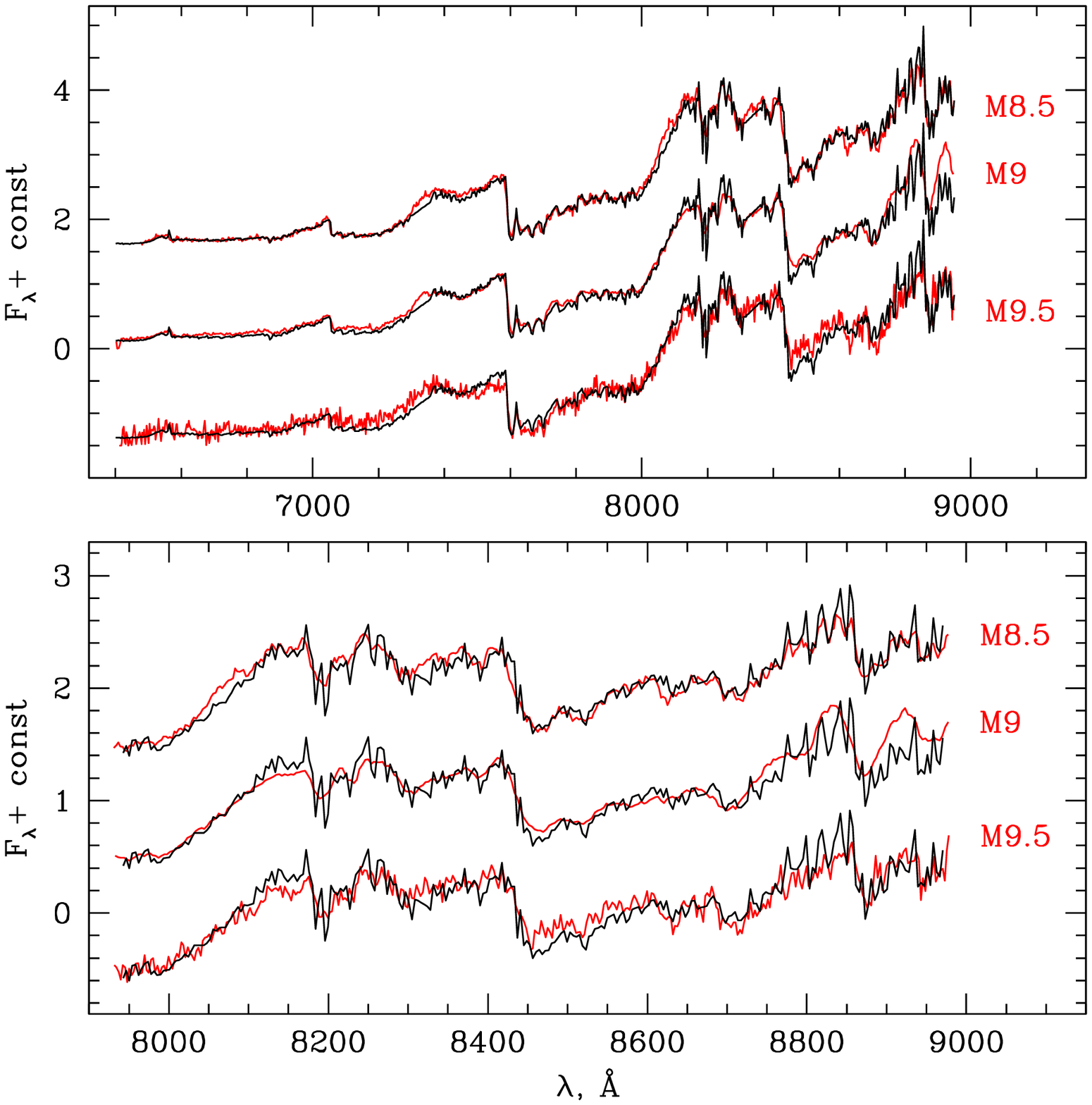}}
\caption{Low (top) and medium (bottom) resolution optical spectra of 
\wise\ (black) compared with template spectra from the Dwarf Archives 
\citep[red; ][]{hen90,kir95,kir99,kir00,kir03,gel04}. Our spectra 
were re-binned by factors of 4 and 12, respectively, for display 
purposes.}\label{fig:spectra_opt}
\end{figure}

A comparison of the overall shape of our optical spectra with 
templates from the Dwarf Archives\footnote{dwarfarchives.org/} 
\citep{hen90,kir95,kir99,kir00,kir03,gel04} suggest a spectral type 
of M8.5--M9 (Fig.\,\ref{fig:spectra_opt}), consistent with the 
estimate of Scholz \cite{sch14}.

\begin{figure}
\resizebox{\hsize}{!}{\includegraphics{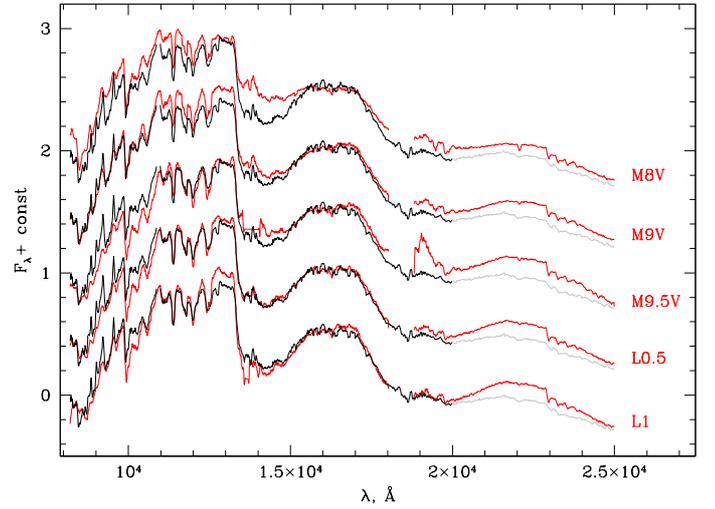}}
\caption{NIR spectrum of \wise\ (black; gray for the region with 
$\lambda$$\geq$20000\,$\AA$ that is not used for spectral typing because 
of the poor wavelength calibration, see Sec.\,\ref{sec:nir_spec}) 
compared with templates from the IRTF library \citep[{red; }][]{cus05,ray09}. 
Similarly to Fig.\,\ref{fig:spectra_all}, the cosmic ray hit region is 
omitted.}\label{fig:spectra_nir}
\end{figure}

We also estimated the spectral type of \wise\ from the spectral 
indices defined by Kirkpatrick et al. (2011) and McLean et al. 
\citet{McLean2003}, listed in Tables\,\ref{table:indices_Kirkpatrick} 
and \ref{table:indices_McLean}, using as reference the sets of 
measurements published in these papers. The optical indices proved of 
limited use because of the considerable scatter and the non-monotonic
behavior of some indices near the M/L transition. The NIR CO index 
was omitted from this analysis, because it shows nearly constant 
strength in the range of spectral types of interest. The two Methane 
indices yield limits, because this feature only gains strength in 
extremely cool objects, later than L8 type.

The H$\alpha$ equivalent width (Table\,\ref{table:eqw}) places \wise\ 
in the regime of the quiescent late M-type field dwarfs 
\citep{Mohanty03}. The lack of Li\,\,{\rm I} absorption makes it 
likely a main sequence star or a relatively old massive BD 
\citep[$>$0.055\,M$_\odot$; e.g. ][]{Burrows97,Baraffe98} that has 
exhausted its Lithium supply.

The overall appearance of our low resolution NIR spectrum resembles
the spectra of objects at the M-L transition
(Fig.\,\ref{fig:spectra_nir}; ignore the problematic red part, as 
discussed in Sec.\,\ref{sec:nir_spec}).
Finally, our medium resolution NIR spectrum shows no Al\,{\rm I} 
doublet at $\lambda$=16724/16756\,$\AA$. This feature is still 
discernible in M9 stars 
\citep[][Figs. 8, 9]{cus05}, indicating an early L-type for \wise. 

Summarizing, the optical spectra yield a range of types from M7 to 
L4, with most probable M9, and the NIR spectra suggest an early L. 
Therefore, we assign to \wise\ a spectral type of L0 (with a 
tentative uncertainty of one subtype), which is in the middle of 
the range of values, determined from all available data, with a 
variety of methods. There may be some discrepancy between the 
optical and the NIR spectra, but we do not consider it strong 
enough to suggest a composite M/L spectrum.

\subsection{Spectral Energy Distribution Fitting}\label{sec:sed}

\begin{figure}
\resizebox{\hsize}{!}{\includegraphics{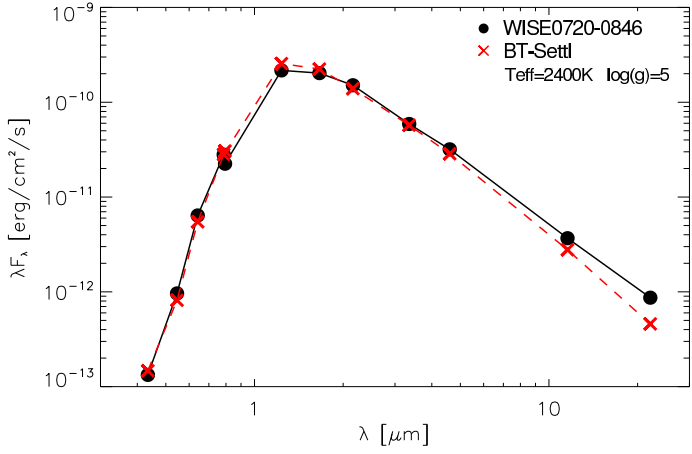}}
\caption{SED fitting of \wise. The black dots show the photometric
measurements, listed in Table\,\ref{table:phot}. The uncertainties 
are comparable to, or smaller than the size of the symbols. For 
details see Sec.\,\ref{sec:sed}}\label{fig:sed}
\end{figure}

The SED of \wise, along with the best-fit BT-Settl \citep{allard11} 
model is shown in Fig.\,\ref{fig:sed}, and magnitudes used are 
listed in Table\,\ref{table:new_HPM_star}. A $\chi^2$ minimization 
was performed with VOSA (Virtual Observatory SED Analyzer; 
\citealt{bayo08}), on a grid of BT-Settl models with effective
temperature T$_{\rm eff}$ between 2000 and 4000\,K, surface gravity 
log\,$g$ between 2 and 6 dex, and metallicity [M/H] between $-$4 
and 0.5. The best fit corresponds to T$_{\rm eff}$=2400\,K, 
log\,$g$=4.5 and [M/H]=0, and it is plotted with red crosses. The 
uncertainties in T$_{\rm eff}$ and metallicity are 100\,K, and 0.5 
dex, respectively, reflecting the spacing of the model grid. The 
VOSA also finds a matching dusty isochrones for the best SED fit 
yielding ages from 0.5 to 1.0\,Gyr and a mass of 
0.06-0.08\,M$_\odot$.

The apparent excess at the WISE 12 and 22\,$\mu$m bands could not 
be accounted for with lower log\,$g$ and T$_{\rm eff}$ -- to obtain 
a satisfactory fit in the MIR, one requires 
T$_{\rm eff}$$\sim$2000\,K, but this would 
predict too low fluxes in the optical pass bands. If real, this 
excess would hint at the presence of a disk, or at an unresolved 
cooler companion. The disk is unlikely given the probable old age 
of \wise. To test the ability of VOSA to match the properties of 
cool objects we also tried fitting LP\,944$-$20 -- the M9 object 
used by Scholz \citet{sch14} as a template. Indeed, the best fit
with BT-Settl models for log\,$g$=5 yields T$_{\rm eff}$=2400\,K, 
and [M/H]=0, leaving again some excess in the WISE bands. The 
excess disappears if log\,$g$=2--3, hinting that LP\,944$-$20 may 
indeed be younger object than \wise, as discussed in 
Sec.\,\ref{sec:intro}.

\subsection{Binarity and a Search for Wide Co-moving Companions}

Scholz \citet{sch14} argued that \wise\ may be a binary because it 
appeared brighter for its spectral type and the parallactic distance
he determined. Some of our data do not support this possibility: our 
improved parallax suggests a distance of 6.07$_{-0.95}^{+1.36}$\,pc, 
in agreement with the apparent brightness of the object, rejecting 
the possibility for nearly equal mass companions; a high mass ratio 
binary is likely to yield conflicting spectral types in the optical 
and in the NIR, which is not the case; a white dwarf companion is 
made unlikely by our $BV$ band imaging and SED fitting, as a simple 
comparison with Sirius B shows; the $K_S$ imaging taken under 
0.6\,arcsec seeing failed to resolve \wise\ into a (wide) binary.

This leaves the possibility for a tight unresolved binary, and the 
difference between the radial velocity derived from the optical and 
NIR spectra seem to support this option. The secondary needs to be 
massive enough. i.e. a BD, on a close in orbit, so it can generate 
a radial velocity amplitude of $\geq$25\,km\,s$^{-1}$ within only 
three days. However, the formal significance of this difference is
at 2$\sigma$ level, and the two spectra were obtained with different 
instruments that are not designed for accurate radial velocity 
monitoring. The disagreement may be due to some hidden systematic 
effects. The question whether \wise\ is a tight binary will need 
additional observations at higher spectral resolution.

We searched for wide co-moving companions to \wise\ by visually 
inspecting 6$\times$6\,arcmin false color images created from 
archival images of the region around \wise, and our NIR images. The 
motion of \wise\ was clearly discernible (Fig.\,\ref{fig:image_PM}). 
The sensitivity of the search was determined by the depth and the 
angular resolution of the shallower image or the image with the worse 
image quality. The most stringent constraints come from the 2MASS--SofI 
comparison: the seeing on the 2MASS images was 2--2.5\,arcsec (to be 
compared with 0.5--0.6\,arcsec on our NIR images); the nominal 2MASS 
completeness limits, determined from the luminosity function downturn, 
for $JHK_S$, are $\sim$15.8, 15.1 and 14.3\,mag, respectively 
\citep{scr06}.

Down to those limits we found no co-moving companions of \wise, but 
we re-discovered another relatively bright ($K_S$=12.8\,mag) 
HPM star 2MASS\,J07200708$-$0845589 (Table\,\ref{table:new_HPM_star}). 
The PPMXL Catalog \citep{Roeser10} reports 
$\mu$$_\alpha$\,cos\,$\delta$=92.7$\pm$7.4, 
$\mu$$_\delta$=$-$167.8$\pm$7.4\,mas\,yr$^{-1}$. 
Adding the positions from WISE and SofI, we obtain:
$\mu$$_\alpha$\,cos\,$\delta$=37.0$\pm$3.8 mas\,yr$^{-1}$, and
$\mu$$_\delta$=$-$124.2$\pm$0.79\,mas\,yr$^{-1}$ with a rather 
large $\chi^2$=356 which is probably due to discrepant WISE and 
SofI positions, probably indicating a different degree of 
background contamination in the NIR and MIR regimens.
The parallax is also poorly constrained -- we find 
$V_{tan}$\,=\,102$\pm$39\,mas,
indicating a distance $d$$\sim$10\,pc. VOSA SED 
fit with the BT-Settl models to CMC15 $r$, DENIS $IJK_S$, 2MASS 
$JHK_S$), and WISE W1 to W3 measurements yielded T$_{\rm eff}$3500\,K, 
log\,$g$=6, and [M/H]=$-$1. The grid spanned the ranges of 
2000--4500\,K, 2--6, and $-$2.5--0, respectively. Weak excess is
visible at W3, but as for \wise, we are reluctant to claim presence
of a disk or a faint companion. Incidentally, there is X-ray source 
1RXS\,J072007.6$-$084541 at 20\,arcsec separation, but Haakonsen \& 
Rutledge \citep{haa09} associate it, albeit with low probability 
for unique association of 0.374, with another objects: 
2MASS\,J07200855$-$0846031.

\begin{table}
\caption{Coordinates of a HPM star 2MASS\,J07200708$-$0845589 in 
the vicinity of \wise. It is heavily contaminated on the WISE 
images by a bright neighbor, and the reddest channel is omitted 
because it is not useful for a position measurement. The 
uncertainties are: 0.45\,arcsec for DSS1, 0.33\,arcsec for DSS2 
Red and IR, and 0.59\,arcsec for DSS2 Blue \citep{mick04}, 
0.06\,arcsec for 2MASS \cite{scr06}, 0.15\,arcsec for WISE 
\citep{wri10}, and 0.015/0.010\,arcsec for SofI, along RA/DEC,
respectively.}\label{table:new_HPM_star} 
\begin{center}
\begin{tabular}{@{ }lcc@{ }}
\hline\hline
Image & RA, DEC (J2000) & Epoch, UT \\
\hline
DSS1 Blue   & 07:20:06.986 $-$08:45:53.13 & 1955-11-19T11:38:00 \\
DSS1 Red    & 07:20:06.995 $-$08:45:52.89 & 1955-11-19T10:43:00 \\
DSS2 IR     & 07:20:07.012 $-$08:45:56.06 & 1981-03-11T11:03:00 \\
DSS2 Blue   & 07:20:07.012 $-$08:45:56.78 & 1982-12-11T16:44:00 \\
DSS2 Red    & 07:20:07.044 $-$08:45:57.43 & 1985-12-15T15:51:00 \\
2MASS $J$   & 07:20:07.086 $-$08:45:59.02 & 1999-02-21T01:27:32 \\
2MASS $H$   & 07:20:07.085 $-$08:45:59.17 & 1999-02-21T01:27:32 \\
2MASS $K_S$ & 07:20:07.089 $-$08:45:59.11 & 1999-02-21T01:27:32 \\
WISE W1     & 07:20:07.108 $-$08:46:00.88 & 2010-04-10T00:15:12 \\
WISE W2     & 07:20:07.098 $-$08:46:00.88 & 2010-04-10T00:15:12 \\
WISE W3     & 07:20:07.072 $-$08:46:01.02 & 2010-04-10T00:15:12 \\
SofI $J$    & 07:20:07.130 $-$08:46:00.73 & 2013-11-14T08:20:41 \\
SofI $K_S$  & 07:20:07.130 $-$08:46:00.93 & 2013-11-14T08:24:44 \\
SofI $H$    & 07:20:07.136 $-$08:46:00.92 & 2013-11-14T08:26:08 \\
\hline
\end{tabular}
\end{center}
\end{table}

\section{Summary and Conclusions}\label{sec:summary}

We carried out a comprehensive follow up observation of \wise, an 
L0 object at $\sim$5-8\,pc from the Sun, reported by Scholz 
\citet{sch14}. New NIR imaging allows us to re-determine the 
distance to \wise, reducing it to  6.07$_{-0.95}^{+1.36}$\,pc, 
bringing into an agreement the geometric and the photometric 
distances -- a binary nature of \wise\ was considered by Scholz 
\citet{sch14} because the object appeared brighter for its spectral 
type and parallactic distance.

Our spectroscopic observations yield spectral type L0$\pm$1, close 
to the M9 determined by Scholz \citet{sch14}. \wise\ is the third 
closest L dwarf to the Sun, after DENIS-P\,J025503.3$-$470049 
\citep[Table 4 in ][]{Kirkpatrick12} and the recently discovered 
WISE\,J104915.57-531906.1A by Luhman \citep{luh13}. For the first 
time we measure the heliocentric radial velocity of \wise, obtaining 
76.56$\pm$2.54 and 101.6$\pm$12.0\,km\,s$^{-1}$ from the optical 
and the NIR spectra, respectively (obtained 3 days apart). They 
agree at 2$\sigma$ level, and cannot be considered an evidence for 
the presence of a close in companion. The weighted average 
heliocentric velocity is V$_{\rm hel}$=77.6$\pm$2.5\,km\,s$^{-1}$. 
Combined with the proper motions, this yields Galactic velocities
that make the object a likely member of the disk or the thick disk.
It is not associated with any of the known nearby young moving 
groups. \wise\ shows weak H$\alpha$ emission, consistent with the 
modest chromospheric activity of field M and L dwarfs. The optical 
spectra show no detectable Li\,\,{\rm I}, suggesting that the 
object is massive enough to have processed its primordial Lithium 
content. 

We also report new $BVRi$ imaging, and use it, together with the 
literature magnitudes to constrain the SED of \wise, yielding 
T$_{\rm eff}$$\sim$2400\,K, consistent with the determined spectral 
type. We note also a weak excess at $\lambda$$>$10\,$\mu$m, but 
consider it rather a problem of the models, than a reliable evidence 
for presence of a disk or a faint red companion. Adaptive optics 
observations or high-precision radial velocity monitoring are needed 
to address these possibilities.

Finally, a HPM star, 2MASS\,J07200708$-$0845589 was re-discovered 
serendipitously in the field of \wise, and we report improved PMs for 
this object. 

During the refereeing process we became aware of the work of 
\citet{bur14}, who obtained M9.5 spectral type for \wise, and measured 
radial velocity of 83.8$\pm$0.3\,km\,s$^{-1}$ and distance of 
6.0$\pm$1.0\,pc, in agreement with our results, within the uncertainties. 
Their AO assisted NIR imaging revealed a faint ($\Delta$$H$$\sim$4.1\,mag)
T5 BD companion candidate at 0.14\,arcsec ($\sim$0.8\,AU) separation that 
is undetectable in our data.


\begin{acknowledgements}
Some observations reported in this paper were obtained with the 
Southern African Large Telescope (SALT). Some observations reported 
in this paper were obtained with the ESO New Technology Telescope 
under program 60.A-9700. All SAAO and SALT co-authors acknowledge the 
support from the National Research Foundation (NRF) of South Africa. 
We acknowledge support by the FONDAP Center for Astrophysics 15010003;  
BASAL CATA Center for Astrophysics and Associated Technologies PFB-06; 
the Ministry for the Economy, Development, and Tourism’s Programa 
Inicativa Científica Milenio through grant IC 12009, awarded to The 
Millennium Institute of Astrophysics (MAS); FONDECYT grants No. 1090213, 
1130196, and 1110326 from CONICYT, and the European Southern Observatory. 
JCB acknowledge support from a Ph.D. Fellowship from CONICYT.
RK acknowledges partial support from FONDECYT through grant 
1130140. 
JB acknowledge support from FONDECYT No. 1120601; 
P.G. thanks STFC for support (grant reference ST/J003697/1).
We have 
also made extensive use of the SIMBAD Database at CDS Strasbourg, of 
the 2MASS, which is a joint project of the University of Massachusetts 
and IPAC/CALTECH, funded by NASA and NSF, and of the VizieR catalogue 
access tool, CDS, Strasbourg, France.
Based on data from CMC15 Data Access Service at CAB (INTA-CSIC).
We thanks the anonymous referee for suggestions that helped to 
improve the paper.
\end{acknowledgements}

%
%

\end{document}